\documentclass[12pt,preprint]{emulateapj}

\usepackage{graphicx}
\usepackage{amsmath}

\begin{document}

\title{Test of the FLRW metric and curvature with strong lens time delays}
\author{Kai Liao$^{1}$, Zhengxiang Li$^{2}$, Guo-Jian Wang$^{2}$, Xi-Long Fan$^{3}$}
\affil{
$^1$ {School of Science, Wuhan University of Technology, Wuhan 430070, China.}\\
$^2$ {Department of Astronomy, Beijing Normal University, Beijing 100875, China.}\\
$^3$ {Department of Physics and Mechanical and Electrical Engineering, Hubei University of Education, Wuhan 430205, China.}
}
\email{liaokai@whut.edu.cn}
\email{xilong.fan@glasgow.ac.uk}

\begin{abstract}
  We present a new model-independent strategy for testing the Friedmann-Lema\^{\i}tre-Robertson-Walker
  metric and constraining cosmic curvature, based on future time delay measurements of strongly lensed
  quasar-elliptical galaxy systems from the Large Synoptic Survey Telescope and supernova observations
  from the Dark Energy Survey. The test only relies on geometric optics. It is independent of
  the energy contents of the universe and the validity of the Einstein equation on cosmological scales.
  The study comprises two levels: testing the FLRW metric through the Distance Sum Rule and
  determining/constraining cosmic curvature. We propose an effective and efficient (redshift)
  evolution model for performing the former test, which allows us to concretely specify the violation criterion for
  the FLRW Distance Sum Rule. If the FLRW metric is consistent with the observations, then, on the second level,
  the cosmic curvature parameter will be constrained to $\sim0.057$ or $\sim0.041$ ($1\sigma$), depending on
  the availability of high-redshift supernovae, much more stringent than
  current model-independent techniques. We also show that the bias in the time delay method might be well controlled,
  leading to robust results. The proposed method is a new independent tool for both testing the fundamental
  assumptions of homogeneity and isotropy in cosmology and for determining cosmic curvature. It is complementary to
  cosmic microwave background plus baryon acoustic oscillation analyses, which normally assume a cosmological model
  with dark energy domination in the late-time universe.
\end{abstract}

\section{Introduction}
Cosmology has been flourishing over the last decades with impressive improvements in the quality and quantity of
astronomical observations. At the basis of most of the remarkable achievements in Cosmology lies the fundamental
assumption that the universe on large scales is homogeneous and isotropic, as described by the
Friedmann-Lema\^{\i}tre-Robertson-Walker (FLRW) metric.

With a substantial amount of new data of unprecedented precision soon to become available, both
from ongoing and forthcoming observational missions, we will be able to test the FLRW metric
directly. The test should be independent of the matter contents in the universe and how they interact
with spacetime geometry, for example, through the Einstein equation. Such a test will either strengthen
our existing understanding or reveal new physics. Therefore, such tests deserve our full attention and it
is important they can be performed once the new data become available. In fact, deviations from FLRW geometry
have been proposed theoretically, and can provide an alternative explanation for the infered late-time acceleration
in our universe~\citep{n1,n2,n3,n4,n5,n6,n7}.

Testing the FLRW metric was proposed in~\citet{Clarkson2008}, where comparing observational determinations of
the expansion rate and cosmological distances was suggested. The method was applied
in~\citet{Shafieloo2010, Mortsell2011, Sapone2014}. Another technique using parallax distance and angular diameter
distance was proposed~\citep{Rasanen2014}. Further,~\citet{Rasanen2015} proposed
that the observation of lensing systems, including separations of the images and central velocity dispersions, can be used
in combination with supernova data to test the FLRW metric, in a model-independent way, through the Distance Sum
Rule (DSR). They applied their method to the existing 23 lensing systems and found no violation.
They also obtained a constraint of cosmic curvature: $-1.22<\Omega_k<0.63$ within $2\sigma$ uncertainty.
Note that their method strongly depends on the universal lens models.

Strong lensing has been a powerful tool for both astrophysics and cosmology~\citep{Treu2010}. A typical system consists of a distant quasar,
lensed by a foreground elliptical galaxy, forming multiple images of the AGN and producing time delays due to
geometrical and the Shapiro effects.
The time delay distance, which is a combination of three angular diameters can be extracted from the observed time delay light
curves and high-resolution imaging from the space telescope. It contains information of the spacetime geometry and the properties
of the matter/energy components in the universe.
We find that the time delay distance is similar to the angular diameter
distance ratio, which can be obtained under the assumption of a universal SIS model (and its extensions) of the lens~\citep{Rasanen2015}, suggesting the DSR works in this case as well. In the time delay method, however, this assumption is not employed.
We have therefore performed a detailed study based on the time delay distance.
Note that the test consists of two levels: first, we test the FLRW metric through the DSR and then, if its validity is confirmed, we
can obtain a constraint of cosmic curvature.
There have been many works on constraining the curvature, mainly based on CMB and BAO, suggesting our universe is flat at the per cent
level~\citep{Planck,k1,k2}. However, these constraints are based on assuming specific cosmological models. Model-independent results
have been given in~\citet{Shafieloo2010,Mortsell2011,Sapone2014,Cai2016}, but the constraints were weak as these methods require
constructing the derivative of noisy distance measure data.
Recently, supernovae in combination with Hubble expansion data gave an improved constraint $\Omega_K=-0.140^{+0.161}_{-0.158}$~\citep{Li2016}.

Throughout this paper, we take a flat $\Lambda$CDM universe with matter density $\Omega_M=0.3$ and
Hubble constant $H_0=70~\rm km~s^{-1}~Mpc^{-1}$ as our fiducial model in simulations.
The speed of light $c=1$.

\section{Lensing and supernova observations}
The current number of well-measured time delay lens systems is limited so that we cannot at present obtain an
accurate enough test from existing data. Measuring time delays requires monitoring the light curves for quite a long
time, usually years, with high cadence and long observing seasons. It also relies on accurate algorithms that can deal with
the independent microlensing effects caused by star motions at different image environments.
Once these conditions are met, one can get precise and accurate time delays, see for example the COSMOGRAIL
program~\citep{Tewesa,Tewesb}.

It is inspiring that the upcoming Large Synoptic Survey Telescope (LSST) will find more than 8000 lensed
quasars, some 3000 of which will have well-measured time delays~\citep{OM10}.
A Time Delay Challenge (TDC) program has recently be initiated to test the accuracy of current
algorithms~\citep{Liao2015,Greg2015}.
The average precision of these time delay measurements has been shown to
be $\sim3\%$ through the first challenge (TDC1), comparable with current uncertainty of lens modelling~\citep{S2013}
(see the HOLICOW program~\citep{HOLI,HOLIV}). Considering that the metric Efficiency defined in~\citet{Liao2015} is
$\sim20\%$, the TDC1 gave at least 400 well-measured time delay systems. Note that the TDC1 only simulated light curves
in the $i$ band and information on the images were not given, making the outlier problem severe. The
efficiency is expected be larger, $\sim40\%$ in the LSST 6 band observation including images of the AGNs along
with their hosts.

\begin{figure}
  \includegraphics[width=8cm,angle=0]{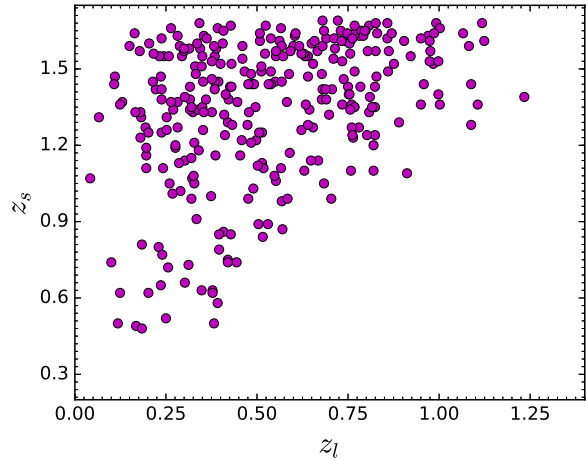}
  \caption{The lens and source redshift distributions of the lensing systems with well-measured time
   delay light curves observed by LSST ($z_s<1.7$).
   The $40\%$ efficiency of the state-of-the-art algorithms would result in $\sim300$ systems.
  }\label{lens}
\end{figure}

In Fig. \ref{lens} we show the redshift distributions of elliptical galaxies and quasars with well-measured time delays
observed by LSST~\citep{OM10}. The source redshift is cut by $1.7$, covering the supernova redshift range.
There are more lensed quasars at higher redshifts, our test is limited by the maximum redshift of the supernovae observations.
Considering the expected improvement of TDC results and the follow-up monitoring programs for existing and
future low redshift source systems as complementary, as well as other projects like the Kunlun Dark Universe
Telescope, the Dark Energy Survey (DES), the South Pole Telescope and so on, we take 100 lensed
quasars with $z_s<1.3$ as a benchmark, which at worse provides an order of magnitude estimate for testing the FLRW
metric and determining cosmic curvature. This number is also consistent with~\citet{Linder2011}.
If the supernova observation can achieve $z\sim1.7$, the corresponding number of lensed quasars would be up to $\sim300$.

The time delay is determined by both the mass distribution of the lens, and a combination of angular
diameter distances known as the time delay distance:
\begin{equation}
D_{\Delta t}=\frac{D_A(z_l)D_A(z_s)}{D_A(z_l,z_s)},
\end{equation}
through
\begin{equation}
\Delta t=(1+z_l)D_{\Delta t}\Delta\phi,
\end{equation}
where $z_l,z_s$ are redshifts at lens and source, respectively.
$\Delta\phi$ is the Fermat potential difference between image positions, which can be inferred
from high resolution imaging observations of the Einstein Ring due to the AGN host galaxy, combined
with spectroscopic observations of the stellar kinematics of the lens galaxy.  The uncertainty on $\Delta\phi$
is expected to be a few percent according to current techniques. Along with the uncertainty on time
delay measurements, it gives $\sim 5\%$ uncertainty on the time delay distance, which is what was used
in~\citet{Linder2011}.

\begin{figure}
  \includegraphics[width=8cm,angle=0]{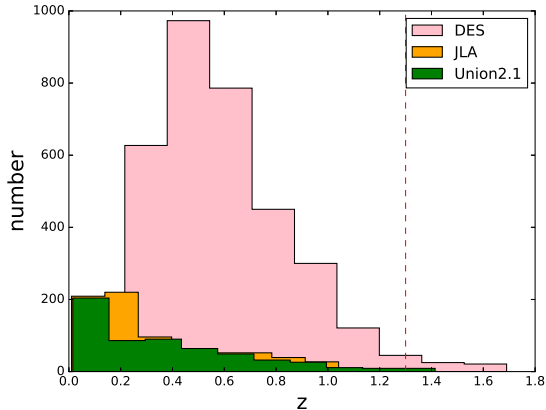}
  \caption{The redshift distributions of supernovae observed by future DES, as well as current Union2.1 and JLA samples. We extend the distribution
  to $z=1.7$ according to the trend of DES.
  }\label{sn}
\end{figure}

We consider supernova observations from DES which carries out a deep optical and near-infrared
survey of $5000\ deg^2$ of the south Galactic cap using a new $3\ deg^2$ CCD camera mounted on the Blanco $4\ m$ telescope
at the Cerro Tololo Inter-American Observatory. It will provide a homogeneous sample of up to 4000 Type Ia supernovae
to better study the nature of dark energy, though the prediction depends on the survey strategy. We take the simulation from
the 10-field hybrid strategy, where the fields are the two deep fields and the three
shallow fields from the 5-field hybrid strategy, plus additional shallow fields clustered around the Chandra Deep Field-South field.
This strategy offers an attractive balance among all important considerations~\citep{Bernstein2012}.
The redshift distribution is shown in Fig. \ref{sn} along with the current Union2.1~
\citep{Union} and JLA samples~\citep{JLA}.
We also expect relatively fewer high-redshift $\sim1.3<z<\sim1.7$ supernovae will be found by future deep supernova surveys,
though the concrete number is not known.
To show the effects of high-redshift supernovae, we just follow the trend of DES prediction made by ~\citet{Bernstein2012} and extend the redshift distribution
to $z=1.7$ in Fig. \ref{sn}, there are $\sim60$ high-redshift supernovae. 

\section{Methodology}
The FLRW metric describes a homogeneous and isotropic universe:
\begin{equation}
d s^2 = - d t^2 + \frac{a(t)^2}{1 - K r^2} d r^2 + a(t)^2 r^2 d\Omega^2,
\end{equation}
where $K$ is a constant describing spatial curvature.
From this metric, the
dimensionless distance $d(z_l,z_s)$ between two redshifts, which is related to the angular diameter
distance $D_A(z_l,z_s)$ through $d(z_l,z_s)= (1+z_s) H_0 D_A(z_l,z_s)$, is given by:

\begin{equation}
d(z_l, z_s)=\frac{1}{\sqrt{|k|}}\begin{cases}  \sinh\left( \sqrt{-k} \chi \right)&k<0\\ \sin\left( \sqrt{k} \chi \right)&k>0,
\end{cases}
\end{equation}
where we denote $k\equiv K/H_0^2=-\Omega_K$ and $\chi=\int_{t_s(z_s)}^{t_l(z_l)} \frac{H_0 d t}{a(t)}$.
Both of these expressions reduce to the flat case in the limit that $k$ goes to zero.
Through the Distance Duality Relation~\citep{Liao2016}, luminosity distance $D_L(z_l,z_s)=D_A(z_l,z_s)(1+z_s)^2$.

Denoting $d(z)\equiv d(0,z)$ as in R$\ddot{a}$s$\ddot{a}$nen et. al. 2015, the DSR that relates $d_{ls}\equiv d(z_l,z_s)$ to
$d_l\equiv d(z_l)$ and $d_s\equiv d(z_s)$ can be written as:
\begin{equation}
d_{ls} = \epsilon_1 d_s \sqrt{1 - k d_l^2} - \epsilon_2 d_l \sqrt{1 - k d_s^2},\label{sumrule}
\end{equation}
where $\epsilon_i=\pm1$, with $k\leq0$ corresponding to $\epsilon_i=1$. If $k>0$, the signs
rely on the locations of the lens and source at the three-dimensional hyper-sphere,
as well as the direction of the light propagation. In this case, the FLRW metric
covers only half of the spacetime. In this work, we assume there exists a one-to-one
correspondence between cosmic time $t$ and redshift $z$, with $d'(z)>0$; then $\epsilon_i=1$.
We rewrite the sum rule as
\begin{equation}
\frac{d_{ls}}{d_ld_s}=T(z_l)-T(z_s),
\end{equation}
where
\begin{equation}
T(z)=\frac{1}{d(z)}\sqrt{1-kd(z)^2}.\label{T}
\end{equation}
Different from Eq.~4 in R$\ddot{a}$s$\ddot{a}$nen et. al. 2015, where the authors relate to the angular diameter distance ratio
obtained from the observed velocity dispersion and image separation, the expression here relates to the
observed time delay distance. Note that the sum rule becomes more symmetric on the right hand side.

To test the FLRW metric, one could use the consistency condition~\citep{Rasanen2015} derived from Eq. \ref{sumrule}:
\begin{equation}
k_S = - \frac{d_l^4 + d_s^4 + d_{ls}^4 - 2 d_l^2 d_s^2 - 2 d_l^2 d_{ls}^2 - 2 d_s^2 d_{ls}^2}{4 d_l^2 d_s^2 d_{ls}^2},
\end{equation}
where the subscript $S$ stands for sum rule.
In this $k_S$ test, the $k_S(z_l,z_s)$ should be a constant equal to $-\Omega_{K0}$ if the FLRW metric is valid.
Any two pairs that give different $k_S$ would indicate a deviation from the FLRW
metric -- a violation of homogeneity and/or isotropy. However, the simulation suggests that the complexity
of the expression may cause non-Gaussian effects which can lead to a bias. Also, the uncertainties
are quire large for individual pairs. In order to get an unbiased estimation, one needs more exquisite statistics, and with current
data quality, the $k_S$ test seems impractical.
A temporary strategy to testing the FLRW metric is to fit a constant $k$ to the data~\citep{Rasanen2015}, in which
case a large $\chi^2/d.o.f$ may indicate a violation of the FLRW metric assumption.

In this paper, we propose an effective and efficient evolution model to test the FLRW metric, where the violation criterion
is specified concretely. As the split term $T$ in Eq.~\ref{T} only relies on a certain redshift in the time delay method, we
extend the constant $k$ to be redshift-dependent $k(z)$:
\begin{equation}
T'(z)=\frac{1}{d(z)}\sqrt{1-k(z)d(z)^2},\label{T1}
\end{equation}
where we simply parameterize $k(z)$ by a first-order Taylor expansion $k(z)=k_0+k_Ez$, the subscript $E$ standing for evolution.
The physical meaning of $k(z)$ deserves further study at the theoretical level. It could be related, for example, to the evolution
of cosmic curvature~\citep{Godlowski2004,Balcerzak2015,Buchert2009}.
In this work, we only focus on observationally testing the validity of the FLRW metric, rather than on the physical motivation of
its alternatives. Therefore, the parameter $k_E$ is used as the violation criterion. Any deviation of $k_E=0$ is a sign of violation
of the homogeneity and isotropy assumptions as represented by the FLRW metric ansatz.

On the second level of our test, if the data support the FLRW metric, we can obtain the probability distribution of $k$, i. e.,
we can constrain the value of the cosmic curvature parameter, thereby testing the flatness of the spatial sections of the universe.

\section{Simulations and results}
We simulated 100 and 300 lensed
quasar-elliptical systems, with source redshifts below $1.3$ and below $1.7$ respectively from the OM10 catalogue~\citep{OM10} that provides mock
observations of LSST based on realistic distributions of quasars and elliptical galaxies, as well as the observing condition
of the telescope. We considered a $3\%$ uncertainty in time delay measurements and the same order uncertainty for
lens modelling embodied as $\Delta\phi$. These would result in $\sim5\%$ uncertainty in the time delay distance or its
inverse for individual systems. For $d_l$ and $d_s$, we simulated $3540$ supernovae where $60$ are $z>1.3$ based on DES 10-field hybird strategy.

We obtained $d(z)$ and $k(z)$ (or constant $k$) in model-independent way, through fitting to supernovae and
lensing systems simultaneously.
In particular, we parameterized $d(z)$ as a fourth order polynomial $d(z)=z+a_1z^2+a_2z^3+a_3z^4$.
Increasing the order does not affect the final result given the quality of the simulated data.
Note that under the assumption that the high-redshift supernovae are much fewer,
the form of the high-redshift extension in Fig. \ref{sn} affects little on determining $d(z)$.
The biggest influence comes from the significantly increased number of lensed quasars.

We used the minimization
function in Python to find the set of parameters that corresponds to minimum $\chi^2$. To get an unbiased
estimation, we simulated 60000 realizations of the data with different random seeds and repeated the
minimization process. Note that the cosmic distances rely on Hubble constant $H_0$, we treated $H_0$ as a free parameter and marginalized over it like the coefficients of the polynomial. In principle, to achieve a model-independent result, for supernova observation, one has to take the original measurements of observed magnitude $m$, the
stretch factor $x$ and color parameter $c$ in the distance modulus $\mu=m-M+\alpha x-\beta c$, where
$M$ is the absolute magnitude, $\alpha$, $\beta$ are nuisance parameters related to the well-known
broader-brighter and blue-brighter relationships, respectively. ($M,\alpha,\beta$) should be taken as free parameters.
However, for our simulation, we only focus on the power of this method and ignore the detailed techniques.
Therefore, we take the distance module as the observational quantities along with the uncertainties from
~\citet{Bernstein2012}.

\begin{figure}
  \includegraphics[width=8cm,angle=0]{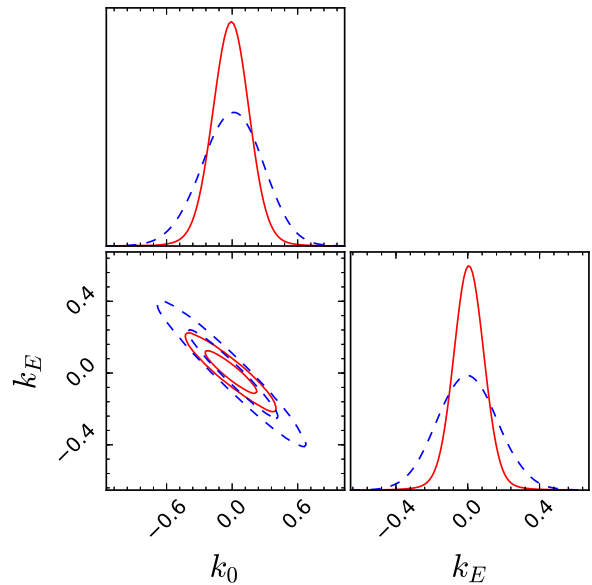}
  \caption{The one-dimensional marginalized distributions and 68\%, 95\% confidence contours from 60000 minimization points
   for the parameters in FLRW test. Two cases are considered where the the maximum redshifts of supernovae are 1.3 (blue line) and 1.7 (red line), respectively.
  }\label{ke}
\end{figure}

\begin{figure}
  \includegraphics[width=8cm,angle=0]{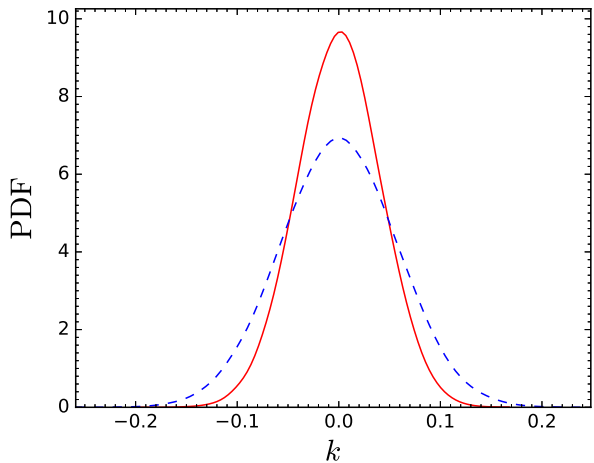}
  \caption{The marginalized distributions of $k$ from 60000 realization minimizations. The colors are consistent with Fig. \ref{ke}.
  }\label{k}
\end{figure}

The marginalized 2-D constraint contour and 1-D probability density distributions of $k_0$ and $k_E$ are shown in Fig.~\ref{ke}.
The marginalized distribution of $k$ is shown in Fig. \ref{k}.
These figures manifestly show that our time delay method is
very powerful on both testing the FLRW metric assumption and constraining cosmic curvature. For cosmic curvature,
the uncertainty is at least one order of magnitude smaller than that in R$\ddot{a}$s$\ddot{a}$nen et. al. 2015. The proposed time delay
method will also surpass the corresponding method using standard clocks~\citep{Li2016} by at least three times.
The numerical uncertainties are presented in Table. \ref{result}.

\begin{table}[!ht]
\large
\begin{center}
\begin{tabular}{lccc}
\hline\hline
$z_{max}$ & $k_0$ & $k_E$ &  $k$\\
\tableline
1.3 &0.271 &0.161 &0.057 \\
1.7 &0.157 & 0.078 &0.041  \\
\hline\hline
\end{tabular}
\end{center}
\caption{The $1\sigma$ uncertainties for the parameters ($k_0$, $k_E$) in FLRW test, as well as the curvature parameter $k$. We consider two cases where
the maximum redshifts of supernovae are 1.3 and 1.7, respectively.
}\label{result}
\end{table}

\section{Summary and discussion}
In this work we developed a new model-independent test of the FLRW metric and cosmic curvature using strong lens
time delay systems and supernovae, based on the Distance Sum Rule. We also introduced an evolution model where
the violation criterion of the FLRW metric is specified. Our method should not be simply deemed as an extension
of R$\ddot{a}$s$\ddot{a}$nen et. al. 2015, where the authors used the angular diameter distance ratio derived from lens velocity
dispersions and image separations.

The distance ratios were derived under the assumption that there exists a simple universal lens model for all lenses, like,
for example, the SIS model or its extensions. In fact, several studies have shown that the universal SIS model (or its extensions)
brings large $\chi^2/d.o.f.$ values, usually $\sim3-4$ in cosmology~\citep{Xia2017,Cao2015,Cao2012}. The reason could be
that the environments of individual systems may be quite different, e.g. due to the effects of nearby galaxies, the densities
of galaxies along the line of sight, or the variation of the lens slopes.
If these can be seen as systematic errors that only enlarge the uncertainty rather than bias the estimation,
one may use the D'Agostini's likelihood~\citep{Xia2017} where a constant intrinsic scatter $\sigma_{int}$, representing any
other unknown uncertainties except for the observational statistical ones, is introduced.
The other way is to simply add an extra Gaussian systematic error to the distance ratio, or the parameter $fe$ that
characterizes the difference between the velocity dispersion of the observed stars and the SIS model velocity dispersion
and accounts for other systematic errors~\citep{Kochanek2000}. However, there might be another possibility, namely that
the large $\chi^2/d.o.f.$ is not caused by systematic errors, but is a bias by the model itself, i. e. assuming a simple universal
model for all lenses is invalid. Outliers are easy to occur if one does not take individuals into consideration. The dispersion
and separation observations may not be sufficient to remove the outliers. In any case, the current distance ratio technique
requires a detailed bias investigation, otherwise results may not be reliable.

The time delay method, on the other hand, is expected to become an unbiased technique. It focuses on individual lensing systems,
each lens can be given an independent flexible model. The measurements of time delays
and the observation of the AGN images along with their host can produce strong constraints on these individual lens models.
There have been ``blind analyses'' conducted to control the bias and make the accuracy
far smaller than the precision for both time delay measurements~\citep{Liao2015} and lens modelling~\citep{HOLI,blind}.
We emphasize that although lens potentials for time delay lenses have been determined accurately, it is currently hard to predict whether
there is any residual systematics after combining many of such time delay lenses. Therefore, the lens model uncertainty
is still very important issue that we have to take seriously.
Furthermore, the distance ratio relies on $\sigma^2$, the square
giving rise to a large uncertainty, while the time delay method relies on the first power for both $\Delta t$ and $\Delta\phi$.

The simulation shows that our method can test the FLRW
metric effectively and efficiently with an uncertainty for $k_E$ $\sim0.078-0.161$.
Under the validity of the FLRW metric, one could achieve an uncertainty on the cosmic curvature parameter $k$ of $\sim0.041-0.057$
in model-independent way, but quite accurately; indeed much more precise than all current techniques. Therefore, we do
not need priors from the Hubble constant and the CMB like in R$\ddot{a}$s$\ddot{a}$nen et. al. 2015.
The independent knowledge of cosmic curvature will significantly contribute to a better understanding of the
evolution history of the universe and of the nature of dark energy as well as inflationary theory which strongly favours a flat universe.
The time delay method proposed here does not involve taking derivatives of distance
measurements~\citep{Shafieloo2010,Mortsell2011,Sapone2014}, which
introduces large uncertainties. The observational data required for the test will be acquired in the very near future,
while the angular diameter and parallax distance method may suffer from a lack of data~\citep{Rasanen2014}.

\section{Perspectives}
According to the investigation of the LSST observing strategy, a limitation
is the relatively low redshifts of supernovae, compared to a typical
redshift $\sim2-3$ of the lensed quasars observed in LSST. To enhance this method,
one could also utilize more high-redshift luminosity distance sources like gamma ray bursts,
quasars~\citep{quasar} or even gravitational wave observations from standard sirens.
On the other hand, the follow-up monitoring programs and the $1-2\ m$ dedicated
telescopes applied in existing systems where the source redshifts are comparable with supernovae
are encouraged. With more useful lensed quasar systems, the
test would be increasingly improved.

In addition, more kinds of lensing systems are promising in the future, for example, the lensed supernova
whose maximum intensity of the light curve could make contribution to the
time delay measurement. The recent detection of gravitational waves has opened a new window for
astrophysics and cosmology. The time delay of lensed GW with its counterpart, like kilonova/mergenova,
short gamma ray burst or fast radio burst, can be measured quite accurately due to the characteristic
waveform, which may also benefit the proposed test~\citep{Fan2017}.

\section*{Acknowledgments}
We thank A. Avgoustidis for polishing the paper.
K. Liao was supported by the National Natural Science Foundation of China (NSFC) No. 11603015
and the Fundamental Research Funds for the Central Universities (WUT:2017IVB067).
Z. Li was supported by NSFC No. 11505008.
X.-L. Fan was supported by NSFC No. 11673008.

\clearpage

\end{document}